\documentclass[a4paper,fleqn,usenatbib]{mnras}

\usepackage{graphics,epsf}
\usepackage{amsmath}                
\usepackage{amsfonts}               
\usepackage{amssymb}                
\usepackage{epsfig}                 
\usepackage{graphicx}

\usepackage{xcolor}
\definecolor{redak}{rgb}{0.9,0.15,0.05}

\def \kms{~\rm{km~s^{-1}}}

\def \msyr{~\rm{M_{\odot}}~\rm{yr^{-1}}}
\def \cm{~\rm{cm}}

\def \K{~\rm{K}}

\def \AU{~\rm{AU}}

\def \etc{$\eta$~Car~}
\def \days{~\rm{days}}

\def \rmModot{~\rm{M_\odot}}
\def \rmRodot{~\rm{R_\odot}}

\title[Accretion at the periastron passage of $\eta$ Car]{Accretion at the periastron passage of Eta Carinae}

\author[A. Kashi]{Amit Kashi$^{1,2,3}$\thanks{E-mail: \href{mailto:kashi@ariel.ac.il}{kashi@ariel.ac.il}}
\\
$^{1}$Physics Department, Ariel University, Ariel, POB 3, 40700, Israel\\
$^{2}$Physics Department, Technion -- Israel Institute of Technology, Technion City, Haifa 3200003, Israel\\
$^{3}$Minnesota Institute for Astrophysics, University of Minnesota, 116 Church St. SE. Minneapolis, MN 55455, USA\\
}

\date{Accepted 2016 September 07. Received 2016 September 07; in original form 2016 July 21}

\pubyear{2016}

\begin{document}
\label{firstpage}
\pagerange{\pageref{firstpage}--\pageref{lastpage}}
\maketitle

\begin{abstract}
We present high resolution numerical simulations of the colliding wind system $\eta$ Carinae,
showing accretion onto the secondary star close to periastron passage.
Our hydrodynamical simulations include self gravity and radiative cooling.
The smooth stellar winds collide and develop instabilities, mainly the non-linear thin shell instability, and form filaments and clumps.
We find that a few days before periastron passage the dense filaments and clumps flow towards the secondary as a result of its gravitational attraction,
and reach the zone where we inject the secondary wind.
We run our simulations for the conventional stellar masses, $M_1=120 \rmModot$ and $M_2=30 \rmModot$,
and for a high mass model, $M_1=170 \rmModot$ and $M_2=80 \rmModot$, that was proposed to better fit the history of giant eruptions.
As expected, the simulations results show that the accretion processes is more pronounced for a more massive secondary star.
\end{abstract}

\begin{keywords}
accretion, accretion discs --- stars: winds, outflows --- stars: individual ($\eta$ Car) --- binaries: general --- hydrodynamics
\end{keywords}

\section{INTRODUCTION}
\label{sec:intro}

The binary system $\eta$ Carinae is composed of a very massive star, hereafter -- the primary (\citealt{Damineli1996, DavidsonHumphreys1997}) and a
hotter and less luminous evolved main sequence star (hereafter -- the secondary).
The system is unique in several aspects, such as a highly eccentric orbit (\citealt{Daminelietal1997}; \citealt{Smithetal2004}), and strong winds (\citealt{PittardCorcoran2002}; \citealt{Akashietal2006}), that together leads to a strong interaction every 5.54 years during periastron passage, known as the spectroscopic event.
During the event many bands and spectral lines show fast variability (e.g., \citealt{Smithetal2000}; \citealt{DuncanWhite2003}; \citealt{Whitelocketal2004};
\citealt{Stahletal2005}; \citealt{Nielsenetal2007}, \citealt{Daminelietal2008a},\citeyear{Daminelietal2008b}; \citealt{Martinetal2010}; \citealt{Mehneretal2010},\citeyear{Mehneretal2011},\citeyear{Mehneretal2015}; \citealt{Davidson2012}; \citealt{Hamaguchietal2007},\citeyear{Hamaguchietal2016}), and the x-ray intensity drops for a duration of a few weeks (\citealt{Corcoranetal2015} and references therein).
Observations of spectral lines across the 2014.6 event indicate weaker accretion onto the secondary close to periastron passage compared to previous events,
hinting at a decrease in the mass-loss rate from the primary star \citep{Mehneretal2015}.

\cite{Soker2005b} suggested that clumps of size of $>0.1$ per cent the binary separation will be accreted onto the secondary near periastron passages.
Accretion was then used to model the spectroscopic events (\citealt{Akashietal2006}; \citealt{KashiSoker2009a}).
\cite{KashiSoker2009b} performed a more detailed calculation, integrating over time and volume of the density within the Bondi-Hoyle-Lyttleton accretion radius around the secondary,
and found that accretion should take place close to periastron and the secondary should accrete $\sim 2 \times 10^{-6} \rmModot$ each cycle.

Other papers referred to a ``collapse'' of the colliding winds region at the spectroscopic event.
This term remained ambiguous since it was first suggested by  \cite{Daminelietal2008a}, and could
be interpreted either as accretion, shell-ejection event \citep{Falcetaetal2005}, or other possibilities (see \citealt{Teodoroetal2012}).
\cite{Parkinetal2009} did however consider a collapse on to the surface of the secondary star,
and developed a model that gives accretion of $\sim 7 \times 10^{-8} \rmModot$ per cycle.

\cite{Parkinetal2011} performed AMR simulations of the colliding winds, but did not obtain accretion.
However, when performing stationary colliding winds simulations at the time of periastron, their results showed 
unstable wind, mainly as a result of the non-linear thin shell instability \citep{Vishniac1994}, and clumps were formed
and reached up to a very close distance from the secondary.
They also suggested that obtaining clumps that fall towards the secondary is resolution-dependent.

\cite{Akashietal2013} conducted 3D hydrodynamical numerical simulations using the \texttt{VH-1} code to study accretion in $\eta$ Car.
They found that a few days before periastron passage clumps of gas are formed due to
instabilities in the colliding winds structure, and some of these clumps flow towards the secondary.
The clumps came as close as one grid cell from the secondary wind injection zone, implying accretion.
In their simulations, however, although the gravity of the secondary star was included, self-gravity of the wind was not included,
and the resolution was too low to see the accretion itself.

\cite{Maduraetal2013} used SPH simulation to model the colliding winds. Though suggesting that a collapse may occur,
their results never showed any collapse or accretion.
Recent numerical simulation of the periastron passages (e.g., \citealt{Maduraetal2015}; \citealt{Clementeletal2015a}, \citeyear{Clementeletal2015b}) were interested in other aspects, and did not find accretion to take place near periastron passages.

In this work we take a step forward, and use one of the best numerical tools available and run advances simulations
in order to test whether accretion takes place, and to what extent.
In section \ref{sec:simulation} we describe the numerical simulation.
Our results, showing accretion, are presented in section \ref{sec:results}
followed by a summary and discussion in section \ref{sec:summary}.

\section{THE NUMERICAL SIMULATIONS}
 \label{sec:simulation}

We use version 4.3 of the hydrodynamic code \texttt{FLASH}, originally described by \cite{Fryxell2000}.
Our 3D Cartesian grid is extended over $(x,y,z)= \pm 8 \AU$, with the secondary fixed at the center,
orbited by the primary.
Our initial conditions are set $50$ days before periastron.
We place the secondary in the center of the grid and send the primary on a Keplerian orbit orbit with eccentricity $e=0.9$.
We use five levels of refinement with better resolution closer to the center. The length of the smallest cell is $1.18 \times 10^{11} \rm{cm}$ ($\simeq 1.7\rmRodot$). 
This finest resolution covers a sphere of a radius of $\simeq 82 \rmRodot$ centered at $(0,0,0)$.
The next level (half the finest resolution) continues up to a radius of $\simeq 320 \rmRodot$.
This level of resolution covers the apex of the colliding winds from $\simeq 20 \days$ before periastron and on.
As shown below, the instabilities that lead to accretion start only a few days before periastron, 
namely within this level of resolution.
The highest resolution allows to follow in great detail the gas as it reaches the injection zone of the secondary wind and being accreted onto the secondary.
Our resolution here is the same resolution as the detailed periastron simulation of \cite{Parkinetal2011},
but we simulate the periastron passage rather than only stationary stars at periastron. 
To solve the hydrodynamic equations we use the \texttt{FLASH} version of the split PPM solver \citep{ColellaWoodward1984}.

As there are different arguments in the literature regarding the masses of the two stars, we use two sets of stellar masses:
\begin{enumerate}
\item A \emph{conventional mass model}, where the primary and secondary masses are $M_1=120 \rmModot$ and $M_2=30 \rmModot$, respectively \citep{Hillieretal2001}.
\item A \emph{high mass model} with $M_1=170 \rmModot$ and $M_2=80 \rmModot$ (\citealt{KashiSoker2010}, where the model is referred to as the `MTz model'; \citealt{KashiSoker2015})
\end{enumerate}
The orbital period is $P=2023$ days, therefore the semi-major axis is $a=16.64 \AU$ for the conventional mass model, and $a=19.73 \AU$ for the high mass model,.
For both models the stellar radii are taken to be $R_1=180 \rmModot$ and $R_2=20 \rmRodot$.
The stars are being modeled by an approximation of an $n=3$ polytrope that is summed to their respective masses (this is done mainly for visualization purposes).
Gravity is being modeled using the Multigrid Poisson solver.
The mass loss rates and wind velocities are $\dot{M}_1=6 \times 10^{-4} \msyr$, $v_1=500 \kms$
and $\dot{M}_2=10^{-5} \msyr$, $v_2=3\,000 \kms$, respectively.
The wind is being injected radially at its terminal speed from a narrow sphere around each star.
In the process of injecting the winds we neglect the spins of the stars, but the orbital motion of the primary relative to the fixed grid is taken into account. 
For the wind the adiabatic index is set to $\gamma=5/3$.
Our initial conditions at $t=-50 \days$ set the entire grid (except the stars themselves) filled with the smooth undisturbed primary wind.
We let the secondary wind blow for $8$ days while the system is stationary, to allow the secondary wind to propagate and to create the colliding
winds structure on one side, and fill the grid on the other side.
We then let the primary proceed on its Keplerian trajectory around the secondary.
Table \ref{table:parameters} summarizes the values of the model properties we use.
\begin{table*}
\centering
\caption{Parameters we use in our simulations.}
\begin{tabular}{llll} 
\hline
Parameter      & Meaning                      & Conventional                     & High                         \\
               &                              & mass model                       & mass model                   \\
\hline
$P$            & Orbital period               & $2023 \days$                     & $2023 \days$                 \\
$e$            & Eccentricity                 & $0.9$                            & $0.9$                        \\
$a$            & Semi-major axis              & $16.64 \AU$                      & $19.73 \AU$                  \\
$M_1$          & Primary mass                 & $120 \rmModot$                   & $170 \rmModot$               \\
$M_2$          & Secondary mass               & $30 \rmModot$                    & $80 \rmModot$                \\
$R_1$          & Primary radius               & $180 \rmRodot$                   & $180 \rmRodot$               \\
$R_2$          & Secondary radius             & $20 \rmRodot$                    & $20 \rmRodot$                \\
$v_1$          & Primary wind velocity        & $500 \kms$                       & $500 \kms$                   \\
$v_2$          & Secondary wind velocity      & $3\,000 \kms$                    & $3\,000 \kms$                \\
$\dot{M}_1$    & Primary mass loss rate       & $6 \times 10^{-4} \msyr$         & $6 \times 10^{-4} \msyr$     \\
$\dot{M}_2$    & Secondary mass loss rate     & $10^{-5} \msyr$                  & $10^{-5} \msyr$              \\
\hline
\end{tabular}
\label{table:parameters}
\end{table*}

We include radiative cooling based on solar composition from \cite{SutherlandDopita1993}.
The problem with radiative cooling is that it limits the time step to be considerably smaller than the hydrodynamic time step limit imposed by the Courant condition.
The post-shocked primary wind is cooler and denser than the post-shocked secondary wind.
Therefore, its cooling time is much shorter and takes only a few seconds.
At the contact discontinuity the primary and secondary winds mix and the combined properties of the gas are closer to those of the post-shocked primary wind
as it is much denser.
This causes the cells of the contact discontinuity to cool radiatively very fast. Then a new layer of cool and dense gas is formed which in turn
mixes with the next layer of post-shocked secondary wind causing it to cool to few$\times 10^6 \K$, where the cooling function reaches high values.
The numerical effect then propagates until the entire secondary wind cools.
Some codes have developed delicate treatments for this problem (e.g., \citealt{Blondin1994}).
To avoid a runaway numerical cooling we therefore limit the cooling of the gas in a single time step and in each grid cell to be no more than $0.3$ per cent of the thermal energy.
We found that this value allows the post shocked secondary wind to be in a temperature of $\approx 10^8 \K$ as inferred from x-ray observations.
When we changed the limit from $0.3$ per cent to $0.03$ per cent there were no large changes in the resulted wind collision region,
but increasing the value to $3$ per cent resulted in too much numerical cooling.
Neither the cooling time nor the hydrodynamic time are constant in space or time.
Our method is a simplification that allows to run the code in reasonable time steps that still allows different cooling in different locations and times
according to the cooling function.

\section{RESULTS}
\label{sec:results}

The colliding winds structure develops instabilities, even during the $8$~days when the system is stationary. 
The shear flow shows the Kelvin-Helmholtz instability, creating its characteristic waves.
Close to the apex, where the two winds hit each other head-on, the the non-linear thin shell instability is obtained.
The instabilities are not the result of injected large perturbations, but rather seeded by the numerical noise and grow with time.

\begin{figure*}
\centering
\includegraphics[trim = 0.0cm 0.0cm 0.0cm 0.0cm,clip=true,width=0.95\textwidth]{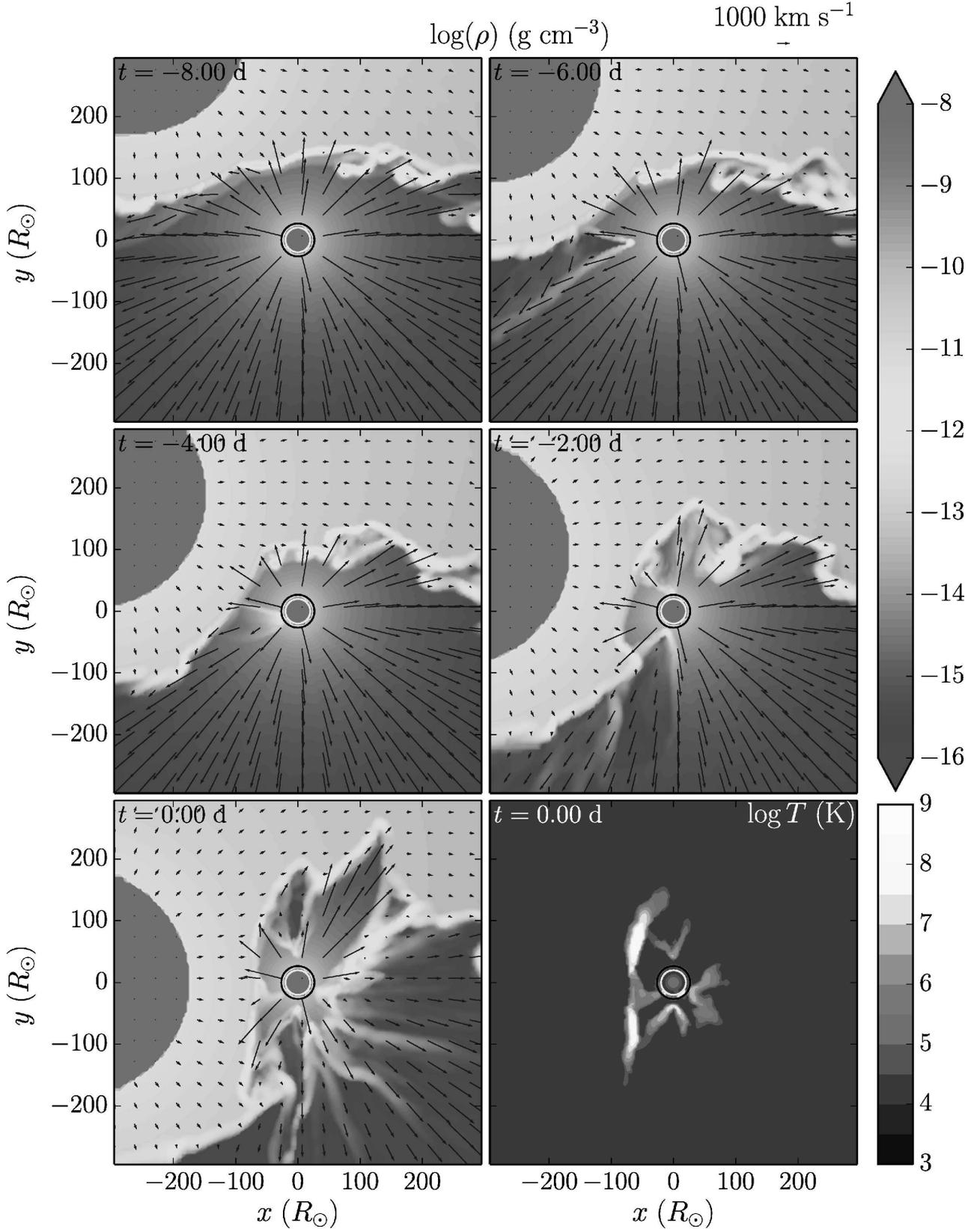}  
\caption{
Density maps showing the density sliced in the orbital plane ($z=0$), for the conventional mass model ($M_1=120 \rmModot$ and $M_2=30 \rmModot$).
The bottom right panel shows a temperature map.
The secondary is at the center, and the primary orbits it from the upper part of the figure to the bottom-left until periastron, and then down-right.
Periastron occurs at $(x,y,z)=(-16.64 \AU, 0, 0)$ and $t=0$.
Times are given with respect to periastron.
The white circle shows the radius of the secondary. The secondary wind is being injected, at terminal velocity, between the white and black circles.
Accretion starts $\approx 4$ days before periastron when the dense clumps that formed in the post-shocked primary wind enter the injection region of the secondary wind.
\textbf{[See the journal for high resolution colour version of this figure.]}
}
\label{fig:density_slices}
\end{figure*}

As the distance between the stars decreases the secondary gravity affects more the colliding winds region.
As can be seen in Figure~\ref{fig:density_slices} the colliding winds region becomes highly unstable \citep{Akashietal2013}, and dense clumps and filaments form.
Some of these clumps and filaments flows towards the secondary.
Some of them enter the injection zone of the secondary wind at $t \approx - 4 \days$, (marked by a black circle in Figure~\ref{fig:density_slices});
this starts the accretion phase.
%

Figure~\ref{fig:density_slices} shows density maps in the orbital plane ($z=0$), for the conventional mass model ($M_1=120 \rmModot$ and $M_2=30 \rmModot$), at different times
of the simulation. Times are given with respect to periastron.
The secondary is at the center of the grid, and the primary orbits it from the upper part of the figure to the bottom-left until periastron, and then bottom-right.
At periastron the primary is exactly to the left of the secondary.
The white circle shows the radius of the secondary.
The secondary wind is being injected between the white and black circles at its terminal velocity.

We take the accretion condition to be that the dense primary wind reaches the injection region of the secondary wind.
It may well be sufficient that the dense primary wind reaches a somewhat larger radius as the acceleration zone of the secondary wind is not modeled here.
According to our condition, accretion starts at $t \approx - 4$ days (4 days before periastron).
The accretion process is expected to substantially disturb the acceleration of the secondary wind.
However, as in this run we do not treat the response of the secondary star to the accreted gas, we just let the injected secondary wind keep pushing the gas away.
For that, after the dense primary wind reaches the acceleration zone of the secondary wind, this run is not an adequate presentation
of the system (we discuss accretion treatment below).
We present the flow after accretion takes place for illustrative purposes.

\begin{figure}
\centering
\includegraphics[trim= 0.0cm 0.0cm 0.0cm 0.0cm,clip=true,width=1.0\columnwidth]{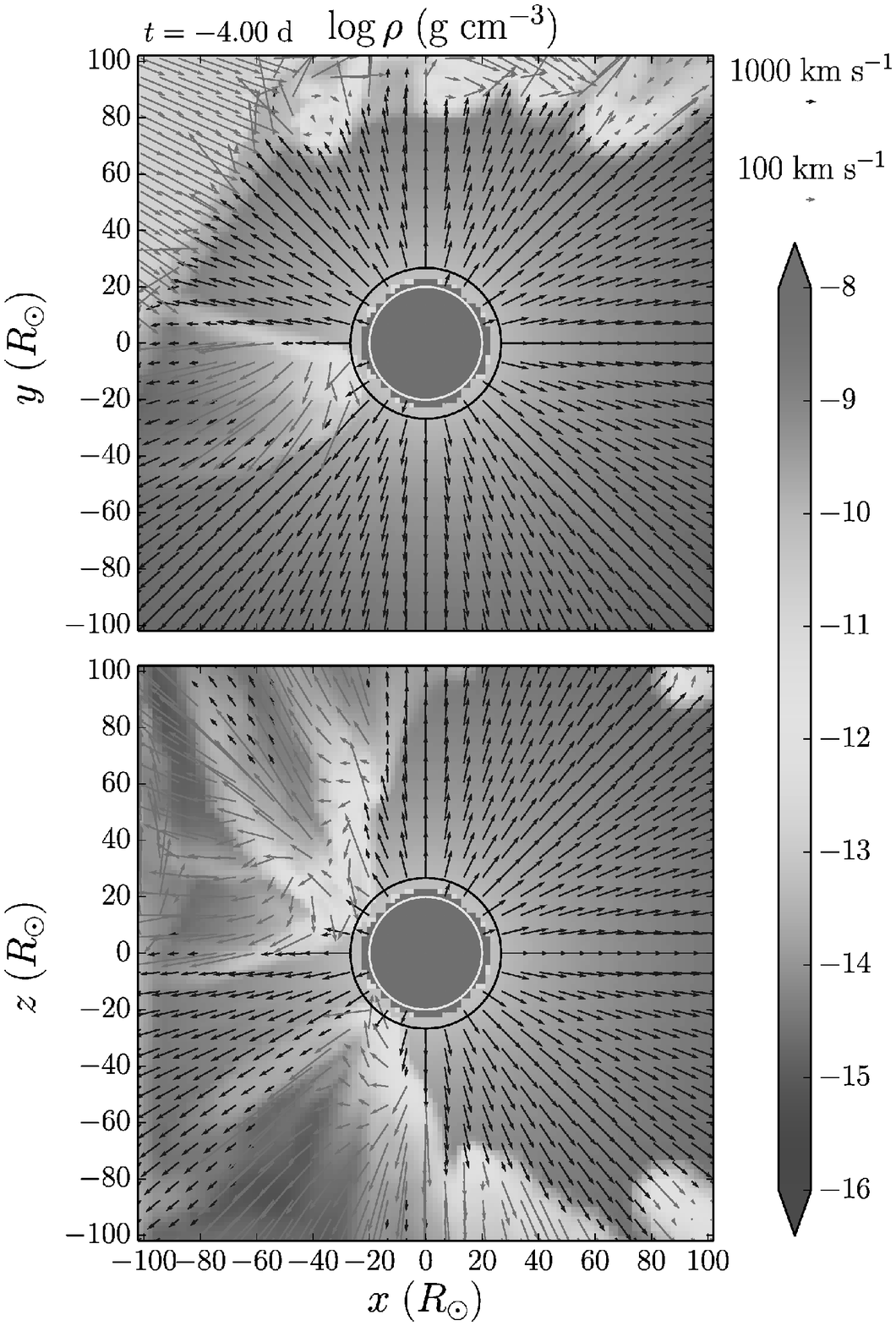}  
\caption{
Zoom-in density maps in the plane $z=0$ (upper panel) ans $y=0$ (lower panel) at $t=-4$~day, for the conventional mass model
($M_1=120 \rmModot$ and $M_2=30 \rmModot$) presented in Figure~\ref{fig:density_slices}.
Zooming-in we can see up to the size of the smallest cells, $1.18 \times 10^{11} \rm{cm}$.
Two scales of velocity vectors are used, black for projected velocities above $1\,000 \kms$, and red for projected velocities below $1\,000 \kms$.
The later are magnified by a factor of 10 for better visibility.
It can be seen that filaments cross the black circle which indicates the outer edge of the wind injection zone, indicating accretion.
\textbf{[See the journal for high resolution colour version of this figure.]}
}
\label{fig:density_slices_zoom}
\end{figure}

\begin{figure*}
\centering
\includegraphics[trim= 0.0cm 0.0cm 0.0cm 0.0cm,clip=true,width=0.49\textwidth]{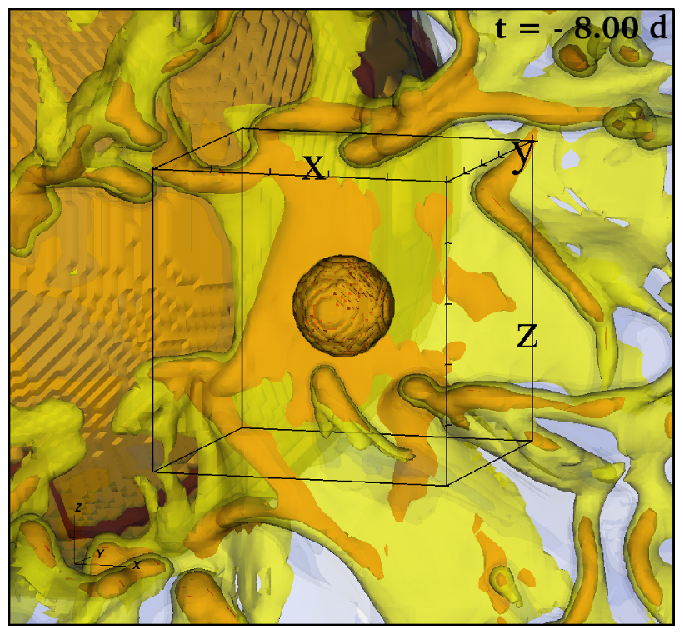}
\includegraphics[trim= 0.0cm 0.0cm 0.0cm 0.0cm,clip=true,width=0.49\textwidth]{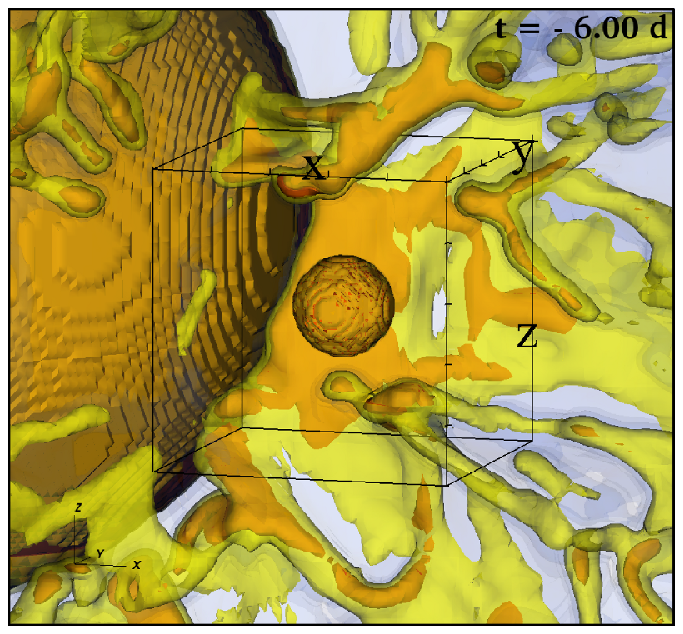}
\includegraphics[trim= 0.0cm 0.0cm 0.0cm 0.0cm,clip=true,width=0.49\textwidth]{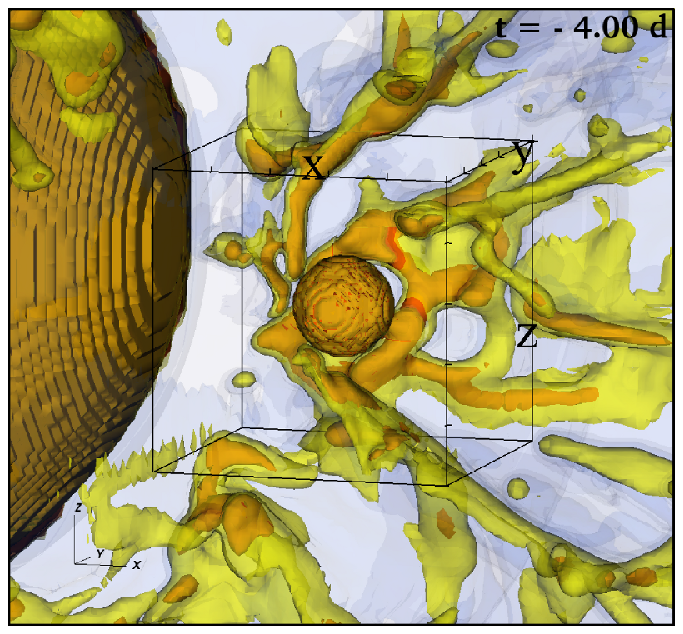}
\includegraphics[trim= 0.0cm 0.0cm 0.0cm 0.0cm,clip=true,width=0.49\textwidth]{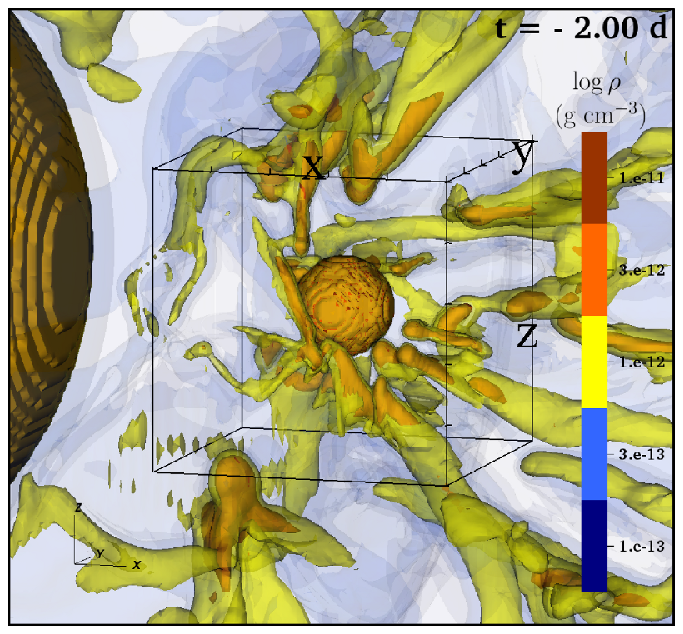}   
\caption{
3D density contours around the secondary for the conventional mass model ($M_1=120 \rmModot$ and $M_2=30 \rmModot$),
taken a few days before periastron passage.
To get a perspective we added a box, centered at the secondary, with side length of $10^{13} \cm$ ($143 \rmRodot$).
The radius of the secondary is $R_2=20 \rmRodot$.
It can be seen that filament of dense gas are formed, and some of them reach very close to the secondary.
The large sphere seen in orange is not the radius of the primary but rather the place where its wind density is at the threshold of
the contour indicated in the legend.
}
\label{fig:density_3d}
\end{figure*}

In Figure~\ref{fig:density_slices_zoom} we show a closer view of the accretion flow structure from two different directions.
It can be seen that filaments flow from different directions.
Figure~\ref{fig:density_3d} shows a 3D view of the accretion for the the conventional mass model at four different times before periastron.
It can be seen that the flow is not at all smooth but rather the shocked primary wind forms many clumps and filaments.
At $t = -8 \days$ the colliding winds structure can still be seen in the back in yellow.
At this time non-linear filaments and clumps are starting to form. The four panels show how the instabilities progress until dense clumps form.
The Plateau–Rayleigh instability might also takes place and create clumps out of the filaments, though our resolution is
not enough to clearly show it.
The secondary wind cannot reverse the inflow of some clumps and filaments, and these are accreted on to the injection zone of the secondary wind. 
%
%
%
\begin{figure}
\centering
\includegraphics[trim= 3.2cm 5.95cm 1.28cm 2.3cm,clip=true,width=0.99\columnwidth]{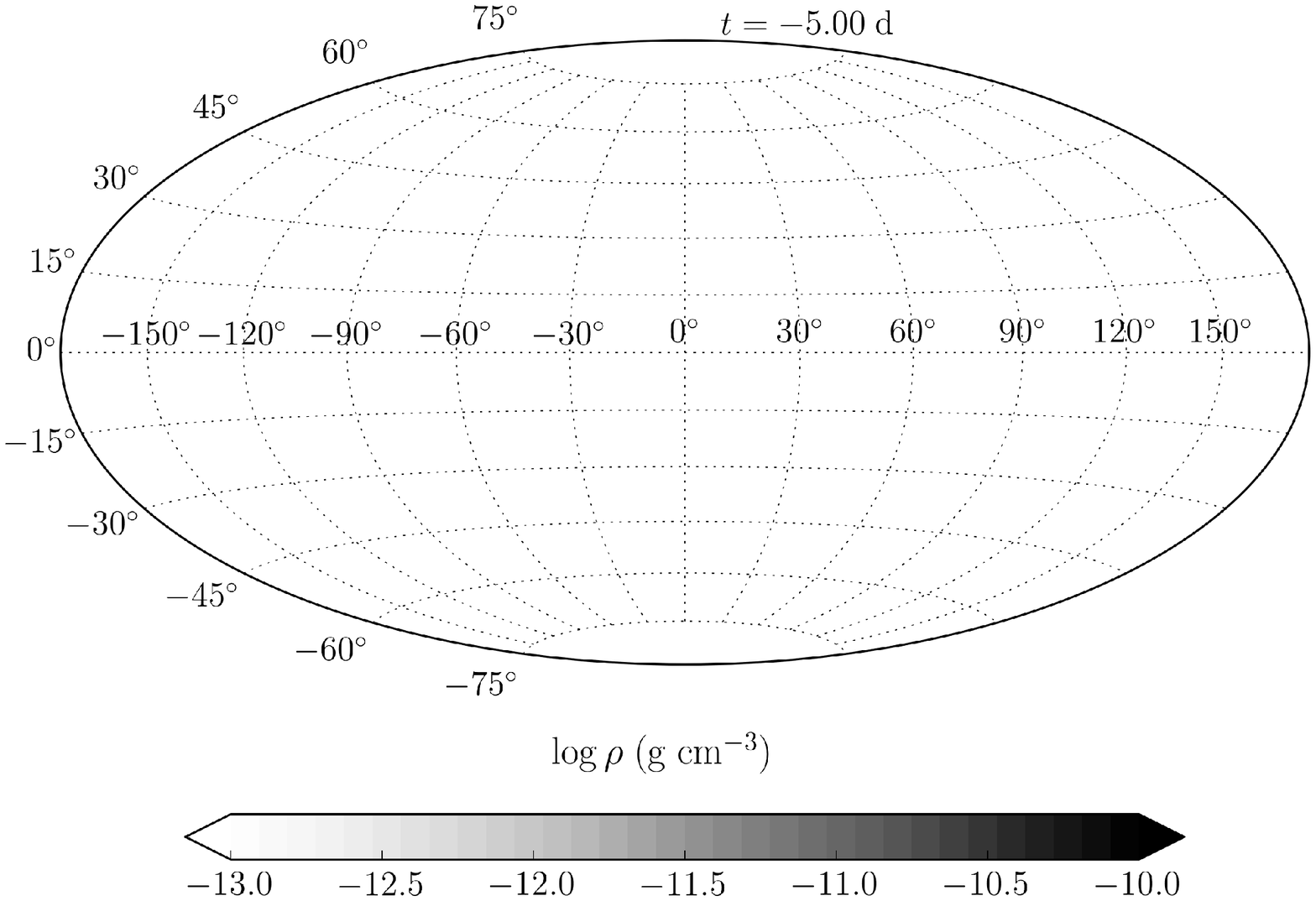}
\includegraphics[trim= 3.2cm 5.95cm 1.28cm 2.3cm,clip=true,width=0.99\columnwidth]{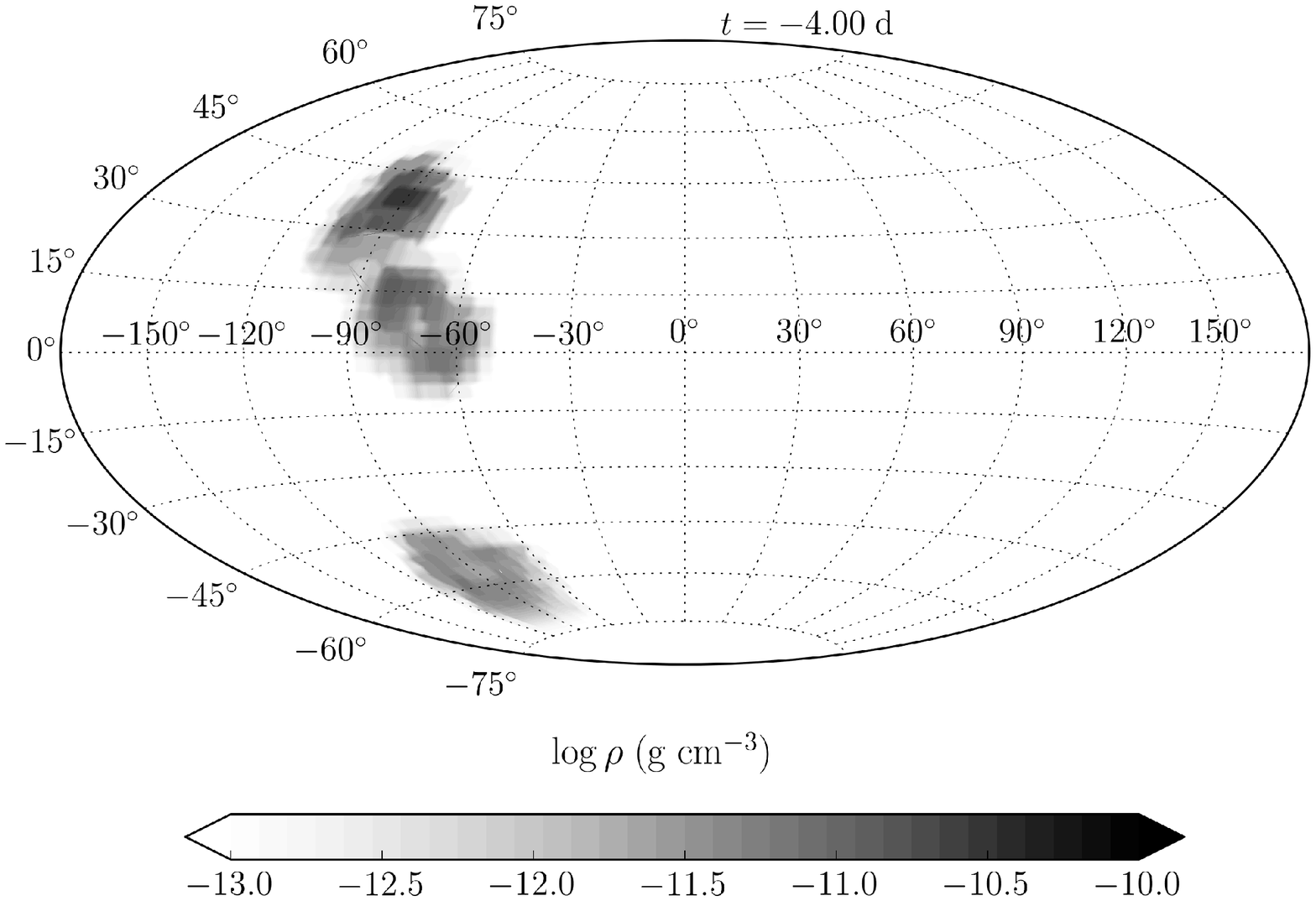}
\includegraphics[trim= 3.2cm 5.95cm 1.28cm 2.3cm,clip=true,width=0.99\columnwidth]{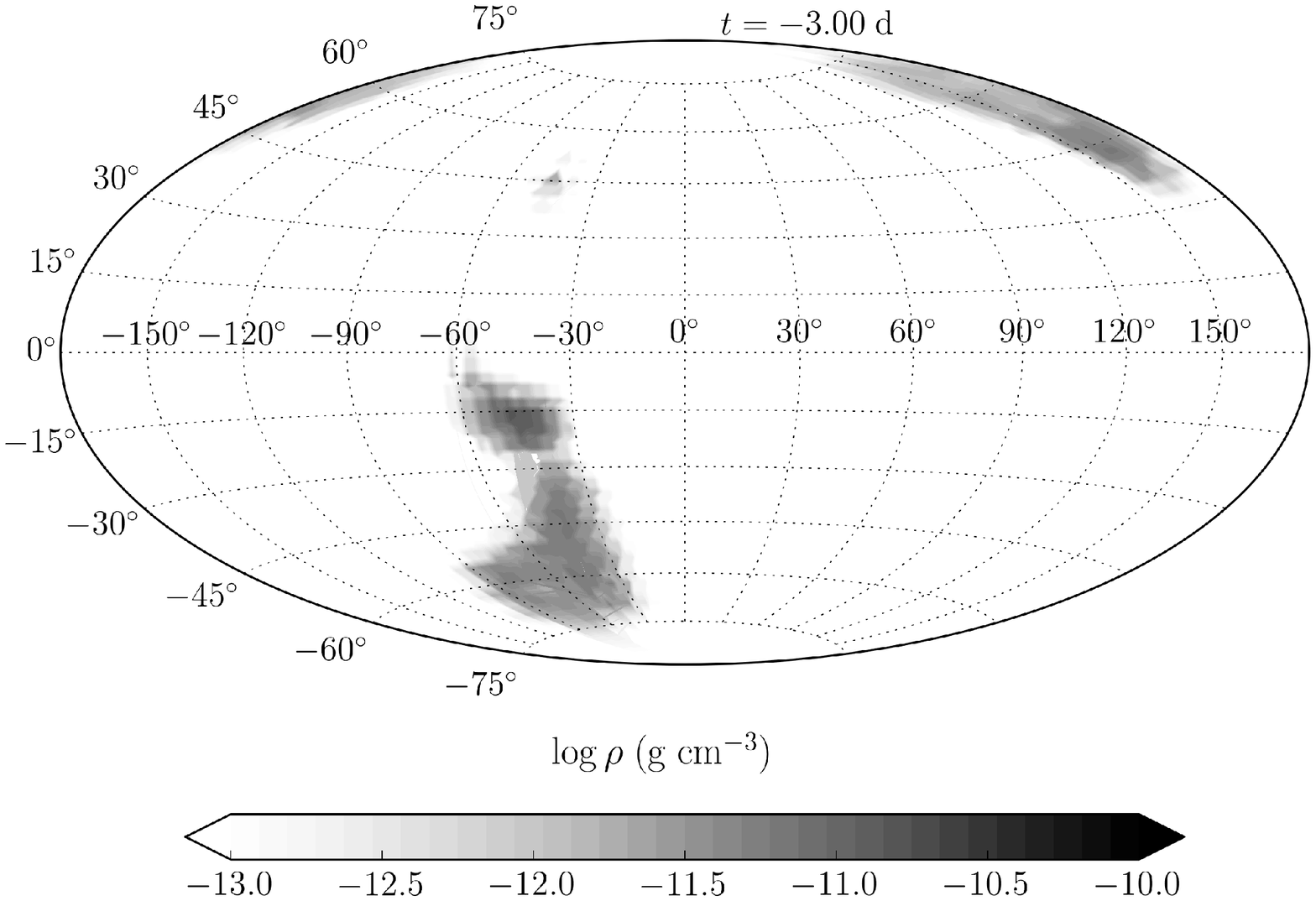}
\includegraphics[trim= 3.2cm 2.65cm 1.28cm 2.3cm,clip=true,width=0.99\columnwidth]{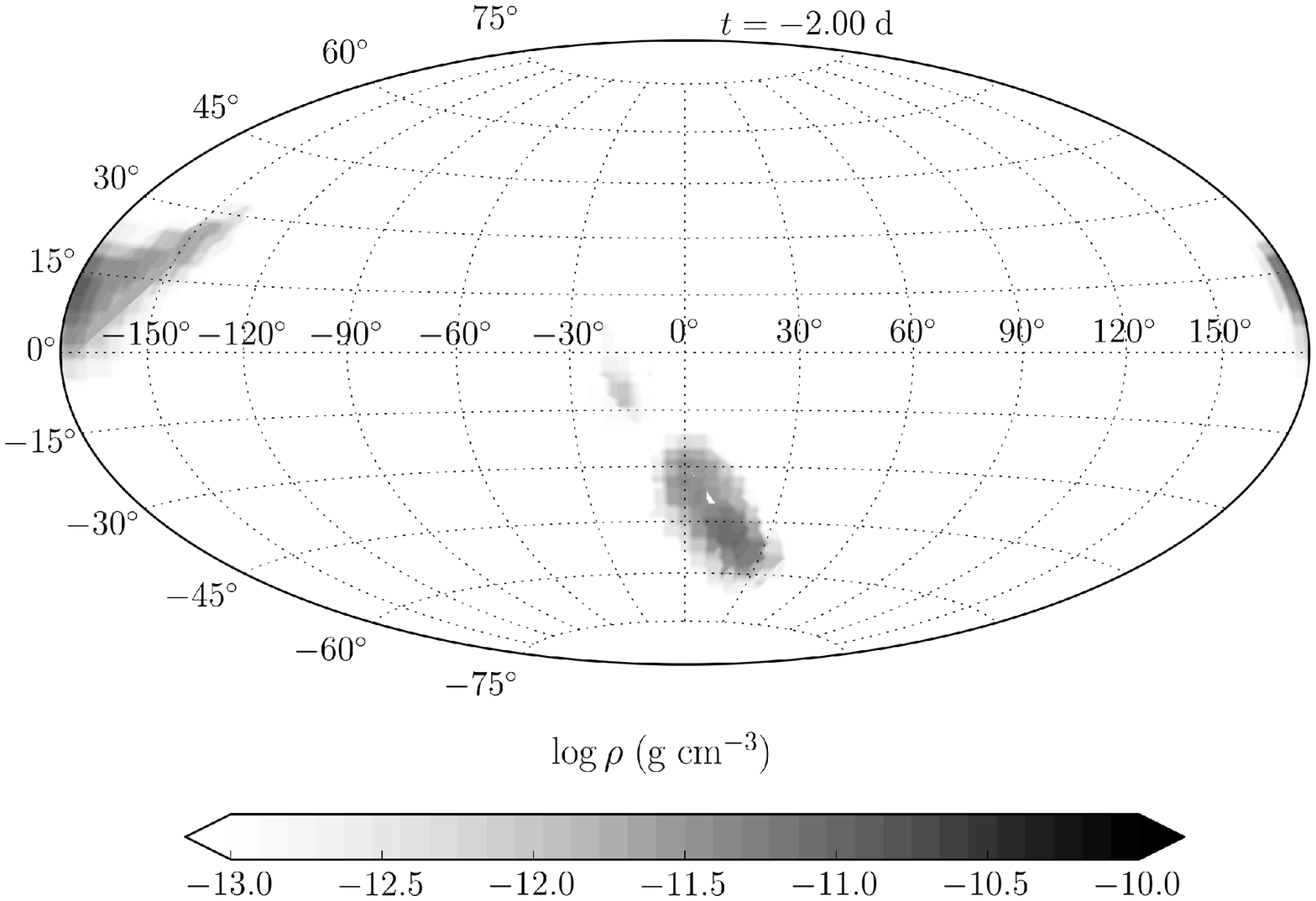}   
\caption{
Density maps on a sphere around the secondary star (Hammer projection) at the edge of the secondary wind ejection region
($r=28 \rmRodot$; black circle in Figure~\ref{fig:density_slices}), for the conventional mass model ($M_1=120 \rmModot$ and $M_2=30 \rmModot$).
The direction $\theta=\phi=0$ points to periastron.
The binary orbit lies on the $\theta=0$ plane (horizontal in the figure).
The white color represents the density of the secondary wind at this radius.
Any darker color (higher density) means that gas is accreted
onto the secondary from that direction.
}
\label{fig:density_spherical}
\end{figure}

\begin{figure*}
\centering
\includegraphics[trim = 0.0cm 0.0cm 0.0cm 0.0cm,clip=true,width=0.95\textwidth]{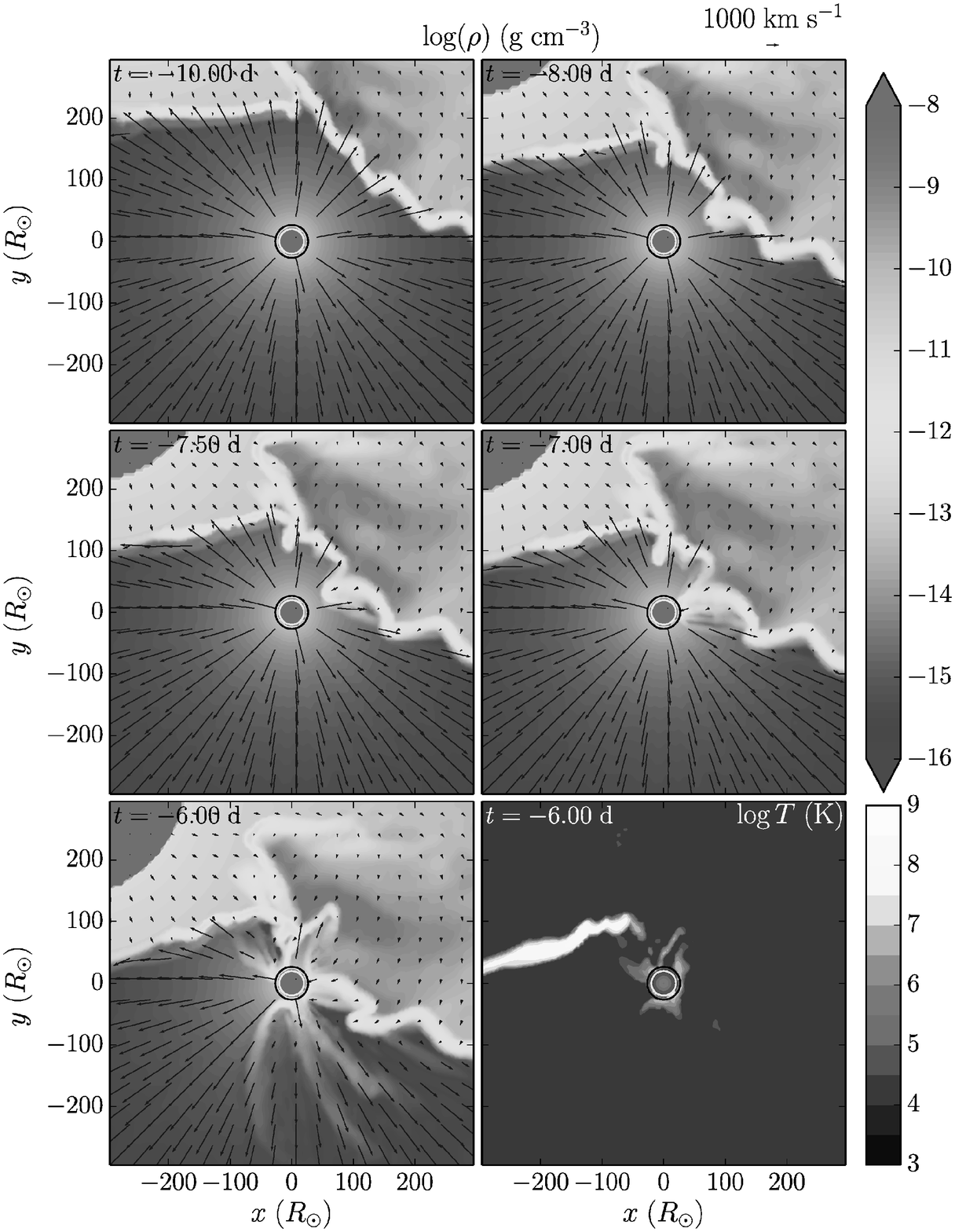}  
\caption{
Like Figure~\ref{fig:density_slices}, but for the high mass model ($M_1=170 \rmModot$ and $M_2=80 \rmModot$),
at selected times close to the beginning of accretion, $\approx 7$ days before periastron.
\textbf{[See the journal for high resolution colour version of this figure.]}
}
\label{fig:density_slices_high_mass}
\end{figure*}

We repeat the simulation for the high mass model (i.e., $M_1=170 \rmModot$ and $M_2=80 \rmModot$), keeping the orbital period and winds properties the same as in the previous simulation. The orbit has the same period and eccentricity, but the semi-major axis is larger, and consequently the periastron distance.
We show the results in Figure~\ref{fig:density_slices_high_mass}.
As the gravitational potential of the secondary is deeper for the high mass model, it more easily attracts the filaments and clumps.
Accretion therefore starts $\approx 7$ days before periastron, about 3 days earlier than for the conventional mass model.
%

An interesting feature obtained in both simulations studied here is the direction of the initial accretion filament.
It could have been anticipated that the accretion would either come from the front, i.e., the direction of the primary,
where material ``climbs'' over the saddle between the potential wells of the two stars, and falls into the potential
well of the secondary, or come from the back, in a Bondi-Hoyle-Lyttleton style. 
Contrary to those expectations, the gas approaches the secondary from the sides, in random directions as can be seen in Figure~\ref{fig:density_spherical}.
This serves as an indication that the clump and filament formation is the dominant process that facilitates the accretion and determines the direction of accretion,
and not only the shape of the gravitational potential or the initial velocities of the flow.

Even though our simulation does not model the reaction of the secondary and its wind to the accreted mass and its likely feedback on the accreted gas,
we measured the amount of mass that reaches the secondary and find it to be in the order of $\approx 10^{-6} \msyr$, in agreement with
the calculations of \cite{KashiSoker2009b}.

\section{SUMMARY AND DISCUSSION}
\label{sec:summary}

We performed detailed \texttt{FLASH} simulations of the \etc colliding winds system close to periastron passage.
The colliding wind region is prone to instabilities that lead to a non-linear formation of clumps and filaments that were accreted onto the secondary star.
The formation of filaments and clumps can occur without self-gravity, as a result of e.g., thermal instability.
The free-fall (collapse) time of each clump as a result of self-gravity is much longer than the duration of the clump formation, indicating self-gravity does not have
a significant role in the formation of the clumps.
Comparing to the simulations of \cite{Akashietal2013} we get a more clumpy flow. This is a result of better resolution and modeling of radiative cooling.
It is important to emphasize that solving the colliding winds and accretion problem requires high resolution to resolve the colliding wind structure and the clumps that
form and then flow towards the accretor (the secondary star in our case).
A delicate treatment of the runaway numerical cooling is essential, otherwise there is a risk of a runaway cooling (see section \ref{sec:simulation}).

The periastron-accretion model advocates for the formation of dense blobs (clumps) in the post-shock primary wind layer of the winds
colliding region \citep{Soker2005a, Soker2005b}.
Based on these early suggestions, there have been a few interpretations of observations of spectral lines accross the periastron passage of \etc as being emitted or absorbed by blobs in the winds colliding region, (e.g, \citealt{KashiSoker2009c}; \citealt{Richardsonetal2016}).
Our simulations confirmed that the dense clumps are crucial to the onset of the accretion process. 

X-ray observations of \etc show flares each cycle when the two stars approach to periastron passage
(\citealt{Davidson2002}; \citealt{Corcoran2005}; \citealt{Corcoranetal2010},\citeyear{Corcoranetal2015}).
\cite{MoffatCorcoran2009} suggested that the flares are the result of that clump formation in the post-shocked primary wind, interacting with the colliding winds region, and compressing the hot gas in the post-shocked secondary wind.

We find that accretion occurs even for smooth primary (and secondary) wind, without creating artificial clumps numerically.
The colliding winds region is compressed in some regions by the instabilities, and that may be the cause for the flares.
Seeding clumps in the primary wind would have also made accretion occur easier.
However, we showed here that even for `rough' conditions -- smooth primary wind and no artificial shut-down of the secondary wind -- accretion does occur.

Accretion is obtained both for the high mass model $(M_1,M_2)=(170 \rmModot, 80 \rmModot)$ and the conventional mass model $(M_1,M_2)=(120 \rmModot, 30 \rmModot)$.
For the high mass model it is easier to obtain accretion, as the stronger secondary gravity attracts the clumps more easily, and does not
let the secondary wind drive them away.

We note that had we turned off the secondary wind in response to clumps reaching the secondary wind injection cells,
accretion rate would have increased, until long after periastron passage.
As the orbital separation increases, accretion rate decreases \citep{KashiSoker2009b}, and the secondary wind is expected to resume. 

Since accretion occurs for the parameters present-day \etc, it let alone occurred during the 1840's Great Eruption, when the mass loss
rate from the primary was larger by orders of magnitude.
The results therefore strengthen the accretion model for the Great Eruption (\citealt{Soker2001,KashiSoker2010}).

When running preliminary further simulations for the conventional mass model, we found that from the beginning of the accretion phase,
up to 50 days after periastron $\approx 10^{-7} \rmModot$
reach the injection zone of the secondary wind.
The amount of mass that is accreted is difficult to estimate, as it requires \emph{modeling} the response of the secondary to the mass that is being accreted.
Specifically, how the secondary wind is affected by accretion.
The number stated above for the accreted mass was derived assuming minimal response.
Obviously, as we know from observations that the x-ray radiation shuts down, this means that the secondary wind is reduced significantly.
We would therefore expect the accreted mass to be larger than the value above, and possibly closer to the results of \cite{KashiSoker2009b}.
In this paper, however, we do not assume anything about the response of the secondary to accretion, and present pure hydrodynamical results
which by themselves show that accretion does occur.

In a future paper we intend to use simulations to model the response of the secondary to the gas accreted onto it.
Doing so will allow us to quantify the accreted mass, and its dependence on the primary mass loss rate and other parameters.
This will hopefully lead to a better understanding of observations of \etc during the spectroscopic event, and the differences between
the last events.

\section*{Acknowledgements}
I thank Kris Davidson and Noam Soker for very helpful discussions and suggestions.
I also thank an anonymous referee for comments that improved the paper.
I acknowledge support provided by National Science Foundation through grant AST-1109394.
This work used computing resources at the University of Minnesota Supercomputing Institute (MSI),
and the Extreme Science and Engineering Discovery Environment (XSEDE),
which is supported by National Science Foundation grant number~ACI-1053575.

\label{lastpage}
\end{document}